\documentstyle[12pt]{article}
\def\@makefnmark{{$\!^{\@thefnmark}$}}
\renewenvironment{thebibliography}[1]
        {\begin{list}{\arabic{enumi}.}
        {\usecounter{enumi}\setlength{\parsep}{0pt}
         \setlength{\itemsep}{0pt} 
         \settowidth
        {\labelwidth}{#1.}\sloppy}}{\end{list}}
\begin{document}
\begin{center}
{\bf Dual Algebraic Pairs and Polynomial Lie Algebras in Quantum
Physics: Foundations and Geometric Aspects}
  \vskip 2em
{V.P. KARASSIOV} \\
{\small\it P.N. Lebedev Physical Institute, Moscow, Russia}\\
\end{center}
\begin{abstract}
\footnotesize
 We discuss some aspects and examples of applications of dual algebraic
pairs  $({\cal G}_1,{\cal G}_2)$ in quantum many-body physics. They arise
in models whose Hamiltonians $H$ have invariance groups $G_i$. Then
one can take ${\cal G}_1 = G_i$ whereas another dual partner ${\cal G}_2=
g^D$ is generated by $G_i$ invariants, possesses a Lie-algebraic structure
and describes dynamic symmetry of models; herewith polynomial Lie algebras
$\hat g = g^D$ appear in models with essentially nonlinear Hamiltonians.
Such an approach leads to a geometrization of model kinematics and dynamics.
\end{abstract}
\section{Introduction}

As is known, group-theoretical and Lie-algebraic methods yield powerful
tools for both qualitative (adequate formulations of model kinematics and
dynamics) and quantitative (dimension reduction of calculations) analysis of
many physical problems $^{1-3}$. In quantum many-body physics, where Hilbert
spaces $L$ of states and all physical observables $O$ are given in terms of
boson ($a_i, a_i^+$) and fermion ($b_j, b_j^+$) operators with standard
commutation relations (CR), Lie-algebraic structures arise in a natural way
via using different boson-fermion mappings:
$(a_i, a_i^+, b_j, b_j^+)\,\stackrel{bfm}\longmapsto\,
F_{\alpha} = F_{\alpha}(a_i, a_i^+,b_j, b_j^+)$
which introduce generators $F_{\alpha}$ of finite-dimensional Lie
(super)algebras $g = {\rm Span}\{F_{\alpha}\}$ as (super)symmetry
operators and simultaneously as {\bf basic} dynamic variables
(i.e. $O=O(\{F_{\alpha}\})$) yielding a most
adequate formulation of problems under study $^3$. Such
algebras  $g $ generate Lie groups $G= \exp  g  = \{\exp F: F \in g\}$
with the key  for applications group property of their elements: $ \exp F_1
\exp F_2 = \exp F_3,\, F_i \in g~^{1,2}$.

Depending on the behaviour of model Hamiltonians $H$ with respect to symmetry
transformations one discerns two (used, as a rule, separately) symmetry
types$~^1$ : a) invariance groups $G_i$ of Hamiltonians $H: \,[G_i, H]_-
\equiv G_i H - H G_i= 0$; b) dynamic symmetry algebras $g^{D}:\, [g^{D}, H]_-
\,\subseteq g^{D} \neq 0 \,( \Longleftrightarrow H \in  g^{D})$. In the first
case Hamiltonians are
considered to be functions in only $G_i$ -invariant (Casimir) operators
$\Lambda_j(G_i)$ whose eigenvalues $\lambda_j $ label  energy levels
$E_{\lambda=[\lambda_j]}$, and  dimensions $d~^{G_i}(\lambda)$ of $G_i$-
irreducible representations (IR) $D^{\lambda}(G_i)$ are equal to the
$E_{\lambda}$- degeneracy multiplicities $\mu(\lambda)$. At the same time
algebras $g^{D}$ already generate total spectra $\{E_{\nu}\}$ of "elementary"
quantum system within fixed IRs $D^{\lambda}(g^D)$ and yield  spectral
decompositions
\begin{equation}\label{5}
L(H)|_{g^D}\,=\, \sum_{\lambda}\,\mu(\lambda)\, L(\lambda),\quad L(\lambda)=
{\rm Span}\{|\lambda; \nu\rangle =D^{\lambda}_\nu(g^D) |\lambda\rangle\}
\end{equation}
of Hilbert spaces $L(H)$ of many-body systems in ($\mu$- multiple)
$g^{D}$-invariant subspaces $L(\lambda)$ generated by actions
of the $g^{D}$ - operators $D^{\lambda}_\nu(g^D)$
on eigenvectors $|\lambda\rangle\in L(H)$ of $g^{D}$-invariant operators
$\Lambda_i$. Subspaces $L(\lambda)$ describe formation of "macroscopic
coherent structures" ($g^{D}$-domains) in $L(H)$  which are stable under the
temporal evolution:  $|\Psi (0)\rangle\,\in L (\lambda)\,\Longrightarrow\,
|\Psi(t)\rangle = U_{H}(t)\,|\Psi (0)\rangle\,\in L (\lambda),\,U_{H}(t) =
\exp (-i t H), H \in  g^{D}$, but a
physical sense of $c$-numbers $\mu, \lambda_j$ in Eq. (\ref{5}) still
remains unclear. At the same time within many-body models with
$G_i$ -invariant Hamiltonians one can reveal deep interrelations between
$G_i$ and $g^D$ symmetries which enable not only to elucidate this sense
but also to formulate an unified invariant-algebraic approach for an
efficient analysis of physical problems in such models $^3$. A natural
formal description of the latter is given in terms of novel mathematical
concepts of dual algebraic pairs (DAP) $^4$ 
incorporating actions of both groups $G_i$ and algebras $g^D$ and
polynomial Lie algebras (PLA)$ ^{5}$ 
arising as $g^D$ in models with essentially nonlinear Hamiltonians $^3$.

The DAP techniques enabled us to elucidate a few non-trivial questions
of quantum physics; however, a number of problems concerning 
applications of PLA is still unsolved $^{3,6}$. 
In this work we briefly discuss these problems and ways of their solution
focusing the main attention on geometric aspects. At first we recapitulate
fundamentals of the DAP and PLA formalism in the context of quantum many-body
physics, restricting ourselves for the sake of simplicity by
the boson case and referring to  $^3$ for a  general discussion. Then we
discuss some aspects of our applications of the DAP techniques in quantum
optics $^{3,6}$ and outline prospects of further studies.

\section{Dual algebraic pairs and polynomial Lie algebras in multiboson
physics: a general analysis}
The notion of DAP extracted  from the vector invariant theory of classical
groups $^7$
by Howe $^4$
is defined in the context of many-boson  systems by\\
{\bf Definition 1}.
Let $a_i=(a_{i\alpha})_{\alpha =1}^m, a_i^ + = (a_i)^\dagger, i=1,\dots,n$
be $n$ pairs of boson vector operators transforming according
to two mutually contragredient fundamental IRs $D^1(G)$  and
 $\bar D^1(G)$  of  a certain group $G$:
\begin{equation} \label{7}
a)\,  a^+_{i\alpha}\quad  \stackrel{D^1(G)} {\longrightarrow}\quad  
\tilde a^+_{i\alpha}\, =\, \sum_{\beta =1}^m  u_{\alpha\beta} a^+_{i\beta},
\qquad
b)\,  a_{i\alpha}\quad  \stackrel{\bar D^1(G)} {\longrightarrow}\quad 
\tilde a_{i\alpha} \,=\, \sum_{\beta =1}^m \bar u_{\alpha\beta} a_{i\beta}.
\end{equation}
Consider the associative algebra ${\cal A}^I_G$  of  vector invariants
of the group $G$ generated by \underline{finite} (according to  the
vector invariant theory $^7$) basis ${\cal B}_{G I} = \{I_j : \, [I_j ,\,G]
= 0\}_{j=1}^{d_{GI}}$
of homogeneous polynomials $I_j  = I_j(a_i, a_i^+)$. Endowing it
by the commuting operation $[I_j, I_j] \equiv [I_i,\, I_j]_- $ one gets
a Lie algebra $ g({\cal A}^I_G)$ with the basis
${\cal B}_{G I}$  and defining CR
\begin{equation} \label{8}
[I_i,\, I_j]\,=\, f_{ij} (\{I_l\})\qquad
(\, [I_a,\, f_{bc}]\,+\,[I_b,\, f_{ca}] \,+\, [I_c,\, f_{ab}] \,=\,0\, )
\end{equation}
where $f_{ij} (\{I_l\})$ are (consistent with the Jacobi identities)
polynomials in $I_l$ stemming from CR for $a_i,a_i^+$ and  the invariant
theory. By the construction two algebraic structures ${\cal G}_1= G$ and
${\cal G}_2= g({\cal A}^I_G)$ commute: $\,[{\cal G}_1,\, {\cal G}_2 ]= 0$
and have a \underline{common} center ${\cal C}({\cal G}_1=G,\,{\cal G}_2=
g({\cal A}^I_G)) = {\cal C}:\,[{\cal C} ,\, {\cal G}_{i=1,2}] = 0$. Then they
are said to form \underline{ DAP} $({\cal G}_1, {\cal G}_2)$ induced by the
$G$-actions (\ref{7}) on $v\equiv v_i = Span\{a^+_i\},\bar v = Span\{ a_i\}.
 \diamondsuit$

The Definition 1 entails  a very important for physical applications\\
{\bf Corollary 1} (sometimes inserted in the DAP definition).  Let
\begin{equation} \label{9}
L(v^{\otimes n}) = Span \{|\{n_{i\beta}\}\rangle  \equiv  \prod_{i,\beta}
(a^+_{i\beta})^{n_{i\beta}} | 0\rangle :\,  a_{i\beta} | 0\rangle =0 \}
\,\equiv\, L_F (nm)
\end{equation}
be the Fock space generated by actions of creation operators $a^+_{i\beta}$
on the vacuum vector $|0 \rangle$ and carrying (due to Eqs. (\ref{7}) and the
${\cal G}_2$ definition) reducible representations of both
structures ${\cal G}_1,\, {\cal G}_2$.  Then there holds the decomposition
\begin{equation}\label{10}
 L(v^{\otimes n})\downarrow_{{\cal G}_1\otimes {\cal G}_2 }=
\sum_{[c_i]} L([c_i]), \quad L([c_i]) = Span \{ D ^{[c_i]}({\cal G}_1)
\otimes D^{[c_i]}({\cal G}_2) |[c_i]\rangle\}
\end{equation}
where $L([c_i])$ are ${\cal G}_1\otimes {\cal G}_2 $-invariant subspaces
labeled by eigenvalues $c_i$ of elements $C_i=C_i(a_i,a_i^+)=\tilde C_i
(I_j)$ of the center ${\cal C}= \{C_i\}$) and generated by joint actions
$D^{[c_i]}({\cal G}_1)\otimes D^{[c_i]}({\cal G}_2)$ of both DAP components
on some reference vectors $|[c_i]\rangle\in L_F(nm)$. $\diamondsuit$

Defining relations (\ref{8}) yield finite-dimensional Lie algebras
$g^0({\cal A}^I_G)={\rm Span}\{I^0_l \}= h$ only if all basic invariants
$I^0_j\in {\cal B}_{G I}$ are quadratic polynomials $I^0_j=F_j(a^+_{i\alpha},
a_{i\alpha})$ that holds, e.g., for groups
$G= O(n), U(n), Sp(2n)$. But in the general case  bases ${\cal B}_{G I}$
contain polynomials $\tilde I_j  =\tilde I_j(a^+_{i\alpha}, a_{i\alpha})=
T_j $ of higher orders which form tensor operators $t={\rm Span}\{T_j\}$ with
respect to $h: [h, t] = t$. Then CR in (\ref{8}) do not close to {\it linear}
combinations of invariants $I_j\in {\cal B}_{G I}$, and repeated commutators
lead to infinite-dimensional Lie algebras $g({\cal A}^I_G)$, generally, not
belonging to well-examined classes of the Kac-Moody algebras $^3$. Therefore,
for physical appications it is useful to consider (retaining Eq. (\ref{10}))
DAP with ${\cal G}_2 ={\cal E}({\cal B}_{G I})$ where ${\cal E}({\cal B}_{GI}
)$ are {\bf defined} as enveloping algebras generated by the bases
${\cal B}_{G I}= h \cup t$ and appropriate specifications of CR (\ref{8}).
Such objects, also appeared in other contexts $^5$, are called as polynomial
deformations of Lie algebras or simply PLA (in view of the absence in the
general case one-to-one correspondences between root systems of PLA and
usual Lie algebras $^{3,6}$).

PLA ${\cal E}({\cal B}_{G I})$ being, by the definition above, specific
($t$-tensor) {\bf extensions} of usual Lie algebras $h$ are also
$G$-invariant subalgebras of the universal enveloping algebra ${\cal U}
(w(nm))$ of the Weyl-Heisenberg algebra $w(nm)={\rm Span}\{a_{i\alpha},
a_{i\alpha}^+\}$. It enables one to specify completely CR (\ref{8}) for
them and to develop their representation theory (unlike the case of
arbitrary PLA $^5$). These constructions are especially simple when
$h$-tensors $t$ consist of two Hermitian conjugated irreducible tensors
$t^{\lambda}= \{T_i^{\lambda}: [T_i^{\lambda}, T_j^{\lambda}]=0\},
t^{\bar\lambda}= (t^{\lambda})^\dagger : t= t^{\lambda} + t^{\bar\lambda}$.
Then CR (\ref{8}) are specified as follows
\begin{equation}\label{11}
a)\, [ h, h ] \,=\, h, \quad b)\,[ h,\, t^{\lambda} ] \,=\, t^{\lambda}, \;
[h,\, t^{\bar\lambda} ]\, =\, t^{\bar\lambda}, \quad  c)\,
[ T_i^{\bar\lambda},\, T_j^\lambda]\,=\, {\cal P}_{ij}(h; r ),\;
r\subset{\cal C}
\end{equation}
where ${\cal P}_{ij}(h; r)$ are polynomials of a fixed degree $s\geq 2$ in
$F_j\in h,R_i\in r$ which are found with the help of the Jacobi identities
and (\ref{11}$b$) from the only polynomial ${\cal P}_{{\bar\lambda}{\lambda}}
(\dots) \equiv {\cal P}^\lambda(\dots)$ (corresponding to "extremal"
components $T_{\bar\lambda}^{\bar\lambda}, T_{\lambda}^{\lambda}$ of tensors
$t^{\bar\lambda}, t^{\lambda}$); the latters, in turn, are determined by
explicit expressions $T_{\bar\lambda}^{\bar\lambda}, T_{\lambda}^{\lambda}
\in {\cal U}(w(nm))$.

So, bases ${\cal B}_{G I}\,=\,h\,\cup\,(t=t^{\lambda} + t^{\bar\lambda})$,
centers $r$ and CR (\ref{11}) define a special (very vast) class of PLA
${\cal E}({\cal B}_{G I})= {\cal E}^{\cal P}_r(h;t^{\lambda})$ as the
second component of the DAP $({\cal G}_1=G,{\cal G}_2 $) connected with
$G$ via the appearance of ${\cal P}, r\subset {\cal C}$ in CR (\ref{11}).
In fact, PLA ${\cal E}^{\cal P}_r(h; t^\lambda)$ can be also examined as
abstract PLA beyond the DAP context that is of interest for finding their
representations not containing in (\ref{10}) (as it is the case for usual
Lie algebras $^1$). As an illustration we consider two examples taken from
physics $^{3,6}$.

 A simplest {\it Example 1} is given by PLA ${\cal E}^{\cal P}_{R_1}(h=
u(1)=\{V_0\}; t^\lambda=v_+^{(1)}=\{V_+\})$ defined by the bases ${\cal B}=
\{V_0, V_+, V_- = V_+^\dagger\}, r=\{ R_1: [R_1, V_a ]=0\}$ and CR
\begin{equation}\label{12}
[V_0, V_\pm ] = \pm V_\pm, \qquad  [V_-, V_+] =
{\cal P} (V_0; R_1) = Q (V_0 +1; R_1) -  Q (V_0; R_1)
\end{equation}
where (extracted from concrete physical models) polynomials $Q(V_0; R_1)$ (of
the degree $s+1$) determine the Casimir operators $C^{\cal E}$ of this PLA:
\begin{equation}\label{13}
C^{\cal E} = V_+ V_- - Q (V_0; R_1), \quad [C^{\cal E}, V_{a}] = 0,\quad
C^{\cal E}|_{L_F(nm)}\equiv 0 (\;\Longleftarrow\;\mbox{Eq.} (\ref{10})).
\end{equation}
The PLA ${\cal E}^{\cal P}_{R_1}(u(1);v_+^{(1)})$ can be
also viewed as polynomial deformations $sl^{\cal P}_{pd}(2)$ of the Lie
algebra $sl (2)={\rm Span}\{Y_0, Y_\pm: [Y_0, Y_\pm]=\pm Y_\pm, [Y_-, Y_+] =
\pm 2Y_0\}$ due to their connection via the generalized Holstein-Primakoff
transformation $^{3,6}$
\begin{equation}\label{14}
Y_0 = V_0-R_0- J,\quad Y_+= V_+ [{\phi(V_0)}]^{-1/2},
 \quad Y_-=(Y_+)^\dagger,  \quad [Y_{\alpha}, R_0]=0=[Y_{\alpha}, J]
 \end{equation}
where $R_0, - J$ are invariant "lowest weight" operators and functions
${\phi(V_0)}$ are determined via polynomials $Q(V_0; R_1)$. Furthermore,
PLA ${\cal E}^{\cal P}_{R_1}(u(1); v_+^{(1)})$ admit two conjugate
realizations by (pseudo)differential operators of one complex variable
$z\in{\bf C}$
$$ V_+=z,\; V_0=z{d}/{dz}+R_0,\; V_-= z^{-1}[C^{\cal E} + Q (z {d}/{dz}+R_0;
R_1)], $$
\begin{equation}\label{15}
V_-={d}/{dz},\; V_0=z{d}/{dz}+R_0,\; V_+ =[C^{\cal E} + Q(z {d}/{dz}+R_0;
R_1)] ({d}/{dz})^{-1}
 \end{equation}
with $Q (z{d}/{dz}+R_0;R_1)= \sum_{k=1}^{s+1}\gamma_k z^k ({d}/{dz})^k $
being determined from (\ref{12}) - (\ref{13})$~^{3,6}$.

{\it Example 2} extends the first one and  is given by
the PLA ${\cal E}^{\cal P}_{R_1}(u(2); v_+^{(2)})$ where
$u(2)= \{E_{ij}: [E_{ij}, E_{kl}] =\delta_{jk} E_{il} - \delta_{il}E_{kj}\}$
is the two-dimensional unitary Lie algebra, and $v_+^{(2)}= \{ V^+_{ij}\}$
is its $2$-nd rank symmetric tensor. All components $V^+_{ij}$ and
$V_{ij}=(V^+_{ij})^\dagger \in v_-^{(2)}= v_+^{(2)\dagger}$ are determined
(via the specifications: $[E_{ij},V^+_{kl}] \equiv ad_{E_{ij}} V^+_{kl}=
\delta_{jk} V^+_{il} +\delta_{jl} V^+_{ki}, [E_{ij}, V_{kl}]= -[E_{ij},
V^+_{kl}]^\dagger$ of CR (\ref{11}$b$)) by $u(2)$ adjoint actions
\begin{eqnarray}\label{16}
2 V^+_{12}= ad_{E_{21}} V^+_{11}, \; 2 V^+_{22}= ad^2_{E_{21}}
V^+_{11},\quad 2 V_{12}= -  ad_{E_{12}} V_{11},\;
 2 V_{22}= - ad^2_{E_{12}} V_{11}
\end{eqnarray}
on the "extremal" components $T_2^{(2)}= V^+_{11}\, (ad_{E_{12}} V^+_{11}=
0= ad^3_{E_{21}} V^+_{11}$) and $T_{\bar 2}^{(2)}= V_{11} \,(ad_{E_{21}}
V_{11}=0= ad^3_{E_{12}} V_{11} $) which together with  $V_0= \frac{1}{2}
E_{11}$ generate PLA $sl^{\cal P}_{pd}(2) \sim \hat E^{\cal P}_{R_1}(u(1);
v_+^{(1)})\subset {\cal E}^{\cal P}_{R_1}(u(2); v_+^{(2)})$ with CR
(\ref{12}). Then, using Eqs. (\ref{16}) and the Jacobi identities we can
calculate all polynomials ${\cal P}_{ij;kl} (\{E_{ij};R_1\}) = [V_{ij},
V^+_{kl}] $ in specfications of CR (\ref{11}c) by the $u(2)$ adjoint actions
on ${\cal P}= {\cal P}_{11;11}(\dots)$, e.g., ${\cal P}_{11;12}(\dots)=
\frac{1}{2}ad_{E_{21}} {\cal P}(\dots)$ etc. Evidently, this procedure of
"lifting" PLA $sl^{\cal P}_{pd}(2)$ to PLA ${\cal E}^{\cal P}_r(h;
t^{\lambda})$ is easily extended on the case of any $h=u(N)+u(M)$ and their
irreducible tensors $t^{\lambda}$; however, generalizations of Eqs.
(\ref{14}), (\ref{15}) are still open problems$~^3$.

And now we outline general features of DAP applications in examining
multiboson models with the Hilbert spaces $L(H)=L_F(nm)$ 
and $G_i$-invariant  Hamiltonians
\begin{eqnarray}\label{17}
H^{n;m}_{G I} = \hbar \left\{\sum_{i,j=1}^n \sum_{\alpha,\beta=1}^m \left[
\omega_{ij}^{\alpha\beta} a_{i\alpha}^+ a_{j\beta} + g^{\alpha\beta}_{ij}
a_{i\alpha }^+ a_{j\beta}^+ + g^{\alpha\beta *}_{ij} a_{i\alpha} a_{j\beta}
\right] +  H^{hd}_{G I}(\{ a_{i\alpha }^+, a_{j\beta}^+\}) \right\}
\end{eqnarray}
where $H^{hd}_{G I}(\dots) = H^{hd\,\dagger}_{G I}(\dots)$ are polynomials of
higher ($\ge 3$) degrees describing essentially nonlinear interactions $^3$.
Then  $H_{G I}\in {\cal E}({\cal B}_{G I}= h \cup t)$  where quadratic terms
in (\ref{17}) belong to $h, \,H^{hd}_{G I} \in t$, and the DAP $({\cal G}_1 =
G_i, {\cal G}_2={\cal E}({\cal B}_{G I}) = g^D)$ naturally arise in such
models. Their use reveals a "synergetic" role of $G_i$-invariance and leads
via the introduction of three types of collective variables related to
$r \subset {\cal C}$ ( \underline{integrals of motions} ), $g^D$
( \underline{"cluster" dynamic} variables ) and $G_i$ (\underline{"hidden"
intrinsic parameters} ) to a geometrization of model kinematics and dynamics
that opens possibilities to apply geometrical methods $^{8-11}$ for their
analysis.

Indeed, the Hamiltonians (\ref{17}) can be reformulated in the $G_i$-
invariant form:
\begin{eqnarray}\label{18}
H^{n;m}_{G I} = H^{n;m}_{G I} (\{I_j\}) = \hbar \left[\, \sum_{j}
\Omega_{j} F_j +\sum_{k} \upsilon_k T_k  + \delta (C_i)\right],
\;   F_j \in h,\; T_k \in t,\; C_i\in r
\end{eqnarray}
(with some of coefficients $\Omega_{j}, \upsilon_k$ being equal to zero),
and the decompositions (\ref{10}) for $L(H)=L_F(nm)$ can be viewed as
specifications of Eq. (\ref{5}) because subspaces $L([c_i])$ have a fibre
bundle structure with fibres $L^{{\cal E}({\cal B}_{G I})}([c_i;\nu]) (\sim
L(\lambda)$ in (\ref{5})) generated by actions $D^{[c_i]}({\cal E}
({\cal B}_{G I}))$ on (labelling the fibre bundle bases) vectors $|[c_i;\nu]
\rangle =D^{[c_i]}(G_i^\nu\in G_i) |[c_i]\rangle$. Herewith dimensions
$d^{G_i}([c_i])$ of the $D^{[c_i]}(G_i)$ IRs are equal to multiplicities
$\mu(\lambda)$ in Eq. (\ref{5}) and  describe degeneracies of {\it all}
energy levels within a given subspace $L([c_i])$.
At the (quasi)classical level of analysis, implemented via generalized
coherent states (CS) $^{2,11}$, the decomposition (\ref{10})
induces the fibre bundle representation
\begin{eqnarray}\label{19}
{\cal M} (H)=\bigcup_{[c_i]} {\cal M}^{[c_i]}(\{\xi^I_a;\zeta^D_b\}),\quad
{\cal M}^{[c_i]}(\{\xi^I_a;\zeta^D_b\}) ={\cal M}_{G_i}^{[c_i]}(\{\xi^I_a\})
\times{\cal M}_{g^D}^{[c_i]}(\{\zeta^D_b\})
\end{eqnarray}
of the model phase spaces ${\cal M}(H)\subseteq {\bf C}^{nm}$ where fibres
${\cal M}^{[c_i]} (\{\xi^I_a;\zeta^D_b\})$ are $G_i \otimes g^D$-invariant
algebraic manifolds (or cell complexes)  determined via dequantizing
subspaces $L([c_i])$ and introducing curvilinear coordinates $\xi^I_a$
and $\zeta^D_b$ related to $G_i$- and $g^D$-generators respectively; herewith
the numbers $c_i$
play the role of topological charges (cf. $^{10}$). In general cases
coordinates $\xi^I_a,\,\zeta^D_b$ are introduced via using so-called
"mean-field approximations" as standard ( "averaging" ) procedures of
dequantizing  quantum problems$~^3$. If $G_i =\exp (g_i)$ and $G^D= \exp(g^D
=h)$ are Lie groups coordinates $\xi^I_a,\,\zeta^D_b$ are associated in a
natural way with parameters of special displacement operators $S_{g_i}
(\{\xi^I_a\})= \exp [\sum \phi_b(\{\xi^I_a\}) g_i^b)], g_i^b\in g_i, \,
S_{h}(\{\zeta^D_b\})=\exp [\sum \varphi_b(\{\zeta^D_a\}) F_b)]$ of groups
$G_i,\, G^D$ which define $G_i\otimes G^D$-orbit-type generalized CS $^{2}$
\begin{eqnarray}\label{20}
|\{\xi^I_a ; \zeta^D_b\} ; \psi_0\rangle\;\equiv\; S_{g_i}(\{\xi^I_a\})\;
S_{h} (\{\zeta^D_b\})\, |\psi_0\rangle,\quad |\psi_0\rangle \in L([c_i])
={\rm Span} \{|[c_i];\nu;\kappa\rangle\}
\end{eqnarray}
on $L([c_i])$ and implement a re-parametrization $|\{\alpha_{i\beta}\}\rangle
= |\{\alpha_{i\beta}(\{\xi^I_a;\zeta^D_b\})\}\rangle$ of the Glauber CS $|
\{\alpha_{i\beta}\}\rangle=  D_{nm}(\{\alpha_{i\beta}\})|0 \rangle=\exp (\sum
[\alpha_{i\beta} a^+_{i\beta} - \alpha^*_{i\beta} a_{i\beta}])|0\rangle$
via the factorization $D_{nm}(\{\alpha_{i\beta}\})=S_{g_i}(\{\xi^I_a\})\,
S_{h} (\{\zeta^D_b\})\, D_{11}(\alpha)\,S_{h}^\dagger(\{\zeta^D_b\})\,
S^\dagger_{g_i}(\{\xi^I_a\})$ of $D_{nm}(\{\alpha_{i\beta}\})^{12}$. However,
direct generalizations of Eqs. (\ref{20}) are less efficient for $g^D=
{\cal E}({\cal B}_{G I})$ because explicit expressions for
matrix elements $\langle [c_i];\nu;\kappa|\exp [ \sum \gamma_b I_b ]|[c_i];
\nu;\kappa\rangle$ are absent.

On the other hand, the introduction of three classes of collective variables
($C_i\in r, I_j\in {\cal E}({\cal B}_{G I}), G_i^\nu\in G_i$) leads to a
dimension reduction of dynamical problems governed by Hamiltonians (\ref{18})
in both Schroedinger and Heisenberg (for dynamic variables $I_j=F_j, T_j$)
pictures. Indeed, the Schroedinger and cluster Heisenberg (for $I_j$)
equations can be written in terms of only variables $C_i, I_j$:
\begin{eqnarray}\label{21}
a)\, i\hbar \frac{d U_{H}(t)}{d t}|\Psi_0\rangle= H\,U_{H}(t)|\Psi_0\rangle,
\qquad b)\, i\hbar \frac{d I_j (t) }{d t}= [ I_j (t), \, H ] =
{\cal L}(\{I_j(t) \})
\end{eqnarray}
where $U_{H}(t)$ is the time-evolution operator induced by $H=H_{G I}$ from
Eq. (\ref{18}) and Eqs. (\ref{21}$b$), in a sense, determine a generalized
dynamics on noncommutative algebraic manifolds ${\cal M}^{C^{\cal E}}
(\{I_i\})=\{ I_j :\tilde C_a (\{I_i\})= C_a^{\cal E}\}$ (see (\ref{13})).
If Hamiltonians (\ref{18}) do not contain operators $T_k \in t$ both Eqs.
(\ref{21}) are solved by group-theoretical methods even for time-dependent
$H_{G I}~^{2,3}:\, U_{H}(t)= \exp (\sum_a\nu_a(t) F_a) = \prod_a
\exp (\eta_a(t) F_a),\;I_j (t) = U_{H}(t) I_j  U^\dagger_{H}(t)= \sum_a
B_a (t) I_j,\,I_j \in {\cal B}_{G I}$ where the second (factorized)
form of $U_{H}(t)$ is more adequate for physical calculations in comparison
with the first one. However, such simple expressions are not valid for
general (even time-independent) Hamiltonians (\ref{18}) due to the absence of
the group property for elements of $\exp[{\cal E}({\cal B}_{G I})]$ and
nonlinearity of ${\cal L}(\{I_j(t) \})$ in Eq. (\ref{21} $b)~^3$. In this
case for $ U_{H}(t),\,I_j (t)$ one can get only "$I_j$-power series"
representations
\begin{equation}\label{22}
U_{H}(t)=\sum_{[k_j]} A^H_{[k_j]} (t) \prod_a I^{k_a}_a \equiv {\cal U}_{H}
(\{I_j\};t),\quad I_j (t) = \sum_{[k_j]} B^j_{[k_j]} (t) \prod_a I^{k_a}_a
\equiv {\cal I}_j (\{I_j\};t)
\end{equation}
where the coefficients $ A^H_{[k_j]} (t), B^j_{[k_j]} (t)$ are determined
from differential-difference equations obtained via the substitution of
Eqs.(\ref{22}) in (\ref{21}) and the use of CR (\ref{11}) $^3$. These
equations define (non-classical) special functions related also with
solutions of differential equations stemming from realizations 
of the type (\ref{15}) for PLA ${\cal E}({\cal B}_{GI})$.

However, at present, simple analytical expressions for these functions are
absent even in the case of simplest PLA $sl^{\cal P}_{pd}(2)~^6$ that
necessitates to separate "principal parts" (or asymptotics) ${\cal U}^0_{H}
(\{I_j\};t),\,{\cal I}^0_j (\{I_j\};t)$ in ${\cal U}_{H}(\{I_j\};t)=
{\cal U}^0_{H}(\{I_j\};t)\{1 + \epsilon ([C_a]) {\cal F}'(t) + \dots\},\,
{\cal I}_j(\{I_j\};t)\,\approx\, {\cal I}^0_j (\{I_j\};t)$ which possess
special (simplifying physical calculations) properties and determine
quasiclassical factors in model dynamics$~^6$. So, e.g., one can take
solutions of classical dynamic equations, obtained via averaging Eqs.
(\ref{21}$b$), as suitable approximations for ${\cal I}^0_j (\{I_j\};t)$.
At the same time asymptotics ${\cal U}^0_{H}(\{I_j\};t)$ can be obtained
from (determined by $g^D$ CS $|[c_i];\nu;\xi \rangle= {\cal S}_{{\cal E}
({\cal B}_{G I})}(\xi)|[c_i];\nu;\rangle \in L^{{\cal E}({\cal B}_{G I})}
([c_i;\nu])$) quasiclassical representations of $U_H (t)$:
\begin{eqnarray}
U_H (t) = \sum_{[c_i]}\int d\mu^{[c_i]}(\xi_0)\int d\mu^{[c_i]}(\xi_t)
\, K_{[c_i]}(\xi_t|\xi_0) \,
\sum_{\nu}|[c_i];\nu;\xi_t\rangle\,\langle [c_i];\nu;\xi_0|
\label{23}
\end{eqnarray}
 where $d\mu^{[c_i]}(\xi_0)$ is a ${\cal E}({\cal B}_{G I})$-
invariant measure on ${\cal M}^{[c_i];\nu}(\xi)\subset {\cal M}(H)$ and the
$\nu$-independent (in view of Eq. (\ref{18})) kernel $K_{[c_i]}(\xi_t|\xi_0)
\equiv \langle [c_i];\nu;\xi_0|U_H (t)|[c_i];\nu;\xi_t\rangle=
\int \exp[i{\hbar}^{-1} S^{[c_i]}(z(t)) ] \prod d\mu^{[c_i]}(z(t))$
has the ${\cal E}({\cal B}_{G I})$-path integral form $^{10,11}$. Its
calculation in the stationary phase approximation$~^{10}$  determines
${\cal U}^0_{H}(\{I_j\};t)$. However, the problem of finding adequate
${\cal S}_{{\cal E}({\cal B}_{G I})}(\xi)$ is not still solved completely.

So, within the DAP framework $G_i$-invariance of $H_{GI}$ classifies states
$|\Psi\rangle \in L(H)$ yielding potential kinematic forms for, generally,
degenerate (with $d^{G_i}([c_i])\neq 1$) $g^D$-domains $L([c_i])$.
Non-degenerate $g^D$-domains with the identical IR $D^{[c_i=0]}(G_i)
\equiv \{I\}$ ($I$ is the operator identity)
describe {\it completely} $G_i$-invariant ($G_i$- scalar)
subsystems having unusual (extremal) physical features while degenerate
$g^D$-domains have "rest" $G_i$ characteristics stipulating an appearance
of critical phenomena in $L([c_i])~^3$. At the same time CS techniques and
associated path integral schemes provide efficient tools to solve dynamical
problems enabling to reveal new cooperative phenomena in $G_i$-invariant
models$~^6$. Furthermore, $G_i$-invariance of $L([c_i])$ allows to examine
on $L([c_i])$ $G_i$-dynamics determined by $g^D$-invariant "intrinsic"
Hamiltonians $H(g^a_i\in g_i =ln G_i)$ with considerng $g^D$-variables as
"dummy" ones$~^{3,12}$.

\section{Dual algebraic pairs in action: applications  in  polarization
and nonlinear quantum optics}
In this Section we demonstrate an efficiency of the DAP concept and
techniques on recent examples of their applications in quantum optics.

The \underline{first example}$~^{3,12}$, manifesting the kinematic
significance of DAP, is due to the gauge $SU(2)$ invariance of free light
fields described by Hamiltonians $H_{fl}$ of the form (\ref{17}) with $m=2,\,
\omega_{ij}^{\alpha\beta}=\omega_i \delta_{ij}\delta_{\alpha\beta},\,
g^{\alpha\beta}_{ij} \equiv 0,\, H^{hd}_{G I} \equiv 0$ and the Hilbert
space $L_F(2n)={\rm Span}\{|\{n_{i\pm}\}\rangle\}$ where $i=1,\dots,n,
\beta =\pm$ label, respectively, spatiotemporal (frequency) and polarization
(in the helicity basis) modes of light. Then, taking $G_i= SU(2)\equiv \{\exp
[\sum_{\gamma=0,\pm} u_{\gamma} P_{\gamma}]:\,P_{0}=\frac12 \sum_i (a^+_{i+}
a_{i+} - a^+_{i-} a_{i-}), P_\pm =\sum_i a^+_{i\pm} a_{i\mp}\}$, we get DAP
$({\cal G}_1=SU(2)=G_i, {\cal G}_2 = so^*(2m) \equiv Span \{E_{ij},\, X_{ij},
\, X^+_{ij} = (X_{ij})^\dagger : \, E_{ij}= \sum_{\beta=\pm} a^+_{i\beta}
a_{j\beta},\, X_{ij} = a_{i+} a_{j-} - a_{i-} a_{j+}\}=g^D=h )$ acting
on $L_F(2n)$. The decomposition (\ref{10}) for $L_F(2n)$ is specified by
determining the "polarization domains"
\begin{equation}\label{24}
L(c_1=p)=Span\{|p;\nu;\kappa\rangle\propto (P_+)^{p+\nu} {\cal D}^p_{\kappa'}
(\{E_{ij}\}) (X^+_{12})^{\kappa_1}|p\rangle,\; |p\rangle= (a^+_{1-})^{2p}
|0\rangle\}
\end{equation}
in $L_F(2n)=\sum L(c_1)$ as eigenspaces of the $SU(2)$ Casimir operator
${\bf P}^2= P_0^2+ \frac12 (P_+P_- +P_-P_+)=C_1\in {\cal C}({\cal G}_1=
SU(2),\,{\cal G}_2= so^*(2m)):{\bf P}^2|p;\nu;\kappa\rangle = c_1(p)|p;\nu;
\kappa\rangle$ whose eigenvalues $c_1(p)= p(p+1),\,p =0,\frac12,1,\dots$
determine values $p$ of the polarization ($P$)-quasispin replacing the
non-gauge-invariant usual spin for light fields.

This decomposition of $L_F(2n)$ provides a new (symmetry) treatment of
polarization structure of light $^{3,12}$ that enabled us to reveal an
unusual ({\it coherent}) sort of unpolarized light ($P$- scalar light) qiven
by states $|0_p\rangle \in L(p=0) =Span\{|p=0;\nu=0;\kappa \neq 0\rangle
\propto \prod (X^+_{ij})^{\kappa_{ij}}|0\rangle\}$ (existing for $L_F(2n),
n\geq 2$) with characteristic property
\begin{equation}\label{25}
P_{\alpha=0,\pm}|0_p\rangle = 0 \quad \Longleftrightarrow\quad
\langle 0_p|P^{a_1}_{1}P^{a_2}_{2}P^{a_0}_{0}|0_p\rangle =0
\quad \forall\, a_1+a_2+a_0\geq 1
\end{equation}
of the "polarization vacuum".  For $n=2$ (when $P_{\alpha}=P_{1 \alpha}+
P_{2\alpha}$) in view of Eq. (\ref{25}) states of $P$-scalar light
generalize so-called Bell states widely used in quantum physics for examining
both fundamental (EPR-paradox, teleportation etc.) and applied (design of
quantum computers, optical communication) problems $^{13}$. Furthermore,
they  give {\bf positive} solutions of the problem of existence of
non-stochastic waves of unpolarized light $^3$ (A. Fresnel, 1821) having
the negative solution in classical optics.

According to general remarks of Section 2 polarization domains $L(p)$ are
dynamically stable under Hamiltonians $H_{G_i\otimes g^D} =  H_{fl} +
H_{so^*(2m)} + H_{SU(2)}$ with
\begin{equation}\label{26}
 H_{fl} = \sum_{i} \omega_{i} E_{ii}, \quad
H_{ so^*(2m)} = \sum_{i\neq j} [\omega_{ij} E_{ij} + g_{ij} X_{ij}
+ g^*_{ij} X^+_{ij}], \quad H_{SU(2)} =\sum_{\alpha}\Omega_{\alpha}
P_{\alpha}
\end{equation}
where $H_{so^*(2m)}$ and $H_{SU(2)}$  determine, respectively, dynamics
of biphoton clusters $X^+_{ij}$ (including their production) and a purely
polarization dynamics. These dynamics are adequately described in terms of
the $SU(2)_p\otimes so^*(2m)$-orbit-type CS of the form (\ref{20}) with
$S_{su(2)}(\xi)=\exp(\xi P_+ - \xi^* P_-),\, S_{so^*(2m)}(\{\zeta^u_b;
\zeta^x_b\})=\exp (\sum [\zeta^u_i E_{ii+1} - \zeta_i^{u *}E_{i+1i} +
\zeta^x_i X^+_{ii+1} -\zeta^{x *}_i X^+_{i+1 i}])$
which, in particular, yield elegant solutions of many quantum problems
(such, e.g., as calculations of geometric phases $^3$, developments of
quantum tomography schemes  and analysis of quantum interference
patterns $^{12}$).

The \underline{second example}$~^{3,6}$, leading to applications of PLA
formalism, is given by models with Hamiltonians $H^{mps}(n;s)= \omega_0 a^+_0
a_0+ H^{n;1}_{G I}$ from (\ref{17}), where $g^{\alpha\beta}_{ij}\equiv 0$ and
\begin{equation}\label{27}
H^{hd}_{G I} = H_I(n;s)=\sum_{1\leq i_1,\dots, i_s\leq n} [ g_{i_1\dots i_s}
a^+_{i_1}\dots a^+_{i_s} a_0 + g^*_{i_1\dots i_s} a_{i_1}\dots a_{i_s}
a^+_0], \quad s \geq 2,
\end{equation}
acting on the Hilbert space $L_F(n+1)={\rm Span}\{|\{n_{i}\}\rangle\}_{i=0,
1}^n$ (the "dummy" label $\beta =1$ is omitted) and describing processes  of
multiphoton scattering. In the case of arbitrary $ g_{i_1\dots i_s}$
Hamiltonians $H^{mps}(n;s)$ have the invariance groups $G_i=C_s\otimes
U_{R_1}(1)$ with both discrete ($C_s=\{e^{i 2\pi k N/s}\}_{k=0,1}^{s-1}, N=
\sum_{i=1}^n E_{ii}, E_{ii} = a^+_i a_i $ ) and continuous ($U_{R_1}(1)=
\{exp (i\phi R_1)\},\,R_1= [N +s E_{00}]/[s+1]$ ) factors. Then  $r= \{R_1\},
\,{\cal B}_{G I} =\{E_{ij}=a^+_i a_j,\;  V^+_{i_1\dots i_s} = a^+_{i_1}\dots
a^+_{i_s}  a_0\in  v_+^{(s)},\; V_{i_1\dots i_s}= a_{i_1}\dots a_{i_s}
a^+_0\in  v_-^{(s)} \}$, where $v_+^{(s)}$ is the $s$- rank symmetric
$u(n)$- tensor, and the DAP $({\cal G}_1=G_i=C_s\otimes U_{R_1}(1),{\cal G}_2
= g^D={\cal E}^{\cal P}_{R_1}(u(n); v_+^{(s)})$) acts on $L_F(n+1)$. In view
of the $G_i$  Abelian nature the decomposition (\ref{10}) for $L_F(n+1)$
contains only non-degenerate  $2j+1$-dimensional  $g^D$-domains
\begin{equation}\label{28}
L([c_1,c_2])=  Span\{|[c_i]; \kappa\rangle
\propto {\cal D}^p_{\kappa'}(\{E_{ij}\}) (V^+_{1\dots 1})^{\kappa_1}
|[c_i]\rangle,\,|[c_i]\rangle =  (a^+_1)^k (a^+_0)^{2j} |0\rangle\}
\end{equation}
where $c_1=k=0,1,\dots, s-1, c_2= 2j=0,1,\dots$ are determined by eigenvalues
of $G_i$-invariant operators. At the same time, in view of CR (\ref{11}),
$G_i$-invariant form (\ref{18}) of the  Hamiltonians $H^{mps}(n;s)$ can be
given by the expressions
\begin{equation}\label{29}
H^{mps}(n;s)=  \hbar \, S_{E} (\xi)  \left[\, \sum_{i,j=1}^n \Omega_{ij}
E_{ij} \, +\, \tilde g V^+_{1\dots 1}\, +\tilde g^* V_{1\dots 1}\, +
+\, \frac{\omega_0}{s} (R_1 - N) \right]  S^\dagger_{E} (\xi)
\end{equation}
($S_{E}(\xi)=\exp \{\sum_{i\geq j}[ \xi_{ij} E_{ij} - \xi^*_{ij}
E_{ji}]\}$) which are most suitable for analyzing Eqs. (\ref{21}).

However, nowadays we can get only (quasi)classical solutions of these
equations, and besides, solely in the case $n=1$ when PLA ${\cal E}^{\cal P}
_{R_1}(u(n); v_+^{(s)})$ is reduced to $su^{\cal P}_{pd}(2)$ defined by Eqs.
(\ref{12})$^{3,6}$. For example, in this case Eqs. (\ref{21}$b$) are
nonlinear analogs
\begin{equation}\label{30}
i\hbar\frac{d V_0}{dt}=\tilde g V_+-g^*V_-,\; i\hbar\frac{d V_+}{dt}=-aV_+
-\tilde g^*{\cal P}(V_0),\; i\hbar\frac{d V_-}{dt}= aV_- +\tilde g
{\cal P}(V_0)
\end{equation}
($V_0=\frac{1}{s+1}[ N - s E_{00}], V_+=V^+_{1\dots 1}, V_-=V_{1\dots 1}$)
of the well-known linear Bloch equations for $su(2)$. In turn, solutions of
 Eqs. (\ref{30}) are equivalent to those of the only equation
\begin{eqnarray}\label{31}
  d^{2} V_0(t) /dt^{2} =  a (H - C) - a^{2}V_{0}(t) +
2\mid\tilde g \mid ^{2} {\cal P}(V_0(t))
\end{eqnarray}
which have in the cluster mean-field approximation ($\langle f(\{V_\alpha\})
\rangle= f(\{\langle V_\alpha \rangle\})$) quasiclassical solutions in terms
of (hyper)elliptic functions $^6$ naturally arising in soliton theories $^{9,
10}$. On other hand, using Eqs. (\ref{14}) in this case  one can transform
linear Hamiltonians (\ref{29}) to an essentially nonlinear form
\begin{eqnarray}\label{32}
H^{mps}(1;s)=  \hbar [\Delta  Y_{0}\, +\,Y_+ g(Y_0) + g^\dagger(Y_0) Y_- \,
+\,\delta(R_1)],\quad g(Y_0)=  \tilde g [{\phi(V_0)}]^{1/2}
\end{eqnarray}
depending on variables $Y_\alpha\in su(2)$ that enabled us to obtain (via
path integral representations (\ref{23}) with using $SU(2)$ CS of the form
(\ref{20})) quasi-classical $SU(2)$ -asymptotics
\begin{equation}\label{33}
{\cal U}^0_{H}(\{Y_\alpha\};t) \; = \;\exp [ \sum_i a_i(t) Y_i ],\quad H=
H^{mps}(1;s)
\end{equation}
of the evolution operators $U_H(t)$ where time-dependent coefficients
$a_i(t)$ are determined through solutions of classical versions of Eqs.
(\ref{31})$^{3,6}$.

\section {Conclusion}
So, we demonstrated natural appearances and an efficiency of DAP and PLA
formalism in examining multiboson models with $G_i$-invariant Hamiltonians.
In conclusion we outline some directions of further studies concerning
physical applications.

They include:\, 1)\, specifications of quasiclassical representations
(\ref{23}) for $U_H (t)$ based on determining  adequate form CS related to
exponentials $Exp(\hat g^{\cal P}({\cal B}_{G I}))$ and on generalizations of
the transformations (\ref{14});\, 2)\, extractions of their "group -like"
asymptotics (extending (\ref{33})) and examinations (in  view of Eqs.
(\ref{15})) of connections  of latters with the Maslov quasiclassical
asymptotics for partial equations in quantum mechanics$^{14}$;\, 3)\,
applications  of geometric methods $^{8,9}$ in analysis (cf. $^{9,10}$) of
nonlinear operator evolution equations of the type (\ref{31}) stemming from
the "cluster" Heisenberg equations (\ref{21}$b$), (\ref{30}) and  their
quasiclassical  approximations (taking into account that Eqs. (\ref{32})
together with transformations ${\cal M}_{g^D}^{[c_i]}(\{\zeta^D_b\})\,
\rightarrow\,  S_{[j]}^{2} (\{\xi_b]\})$ of fibers in  (\ref{19})
into the Bloch spheres describe a geometrization of model dynamics).

\section*{Acknowledgments}
The author is thankful to Professor Z. Rakic for  his attention to this work.
 
\end{document}